\documentclass[aps,prd,onecolumn,eqsecnum,amsmath,nofootinbib,preprintnumbers,superscriptaddress]{revtex4}

\usepackage[dvipsnames]{xcolor}
\usepackage{color,graphicx,float,xcolor}
\usepackage{amsfonts,amssymb,theorem,mathrsfs,times}
\usepackage{bm}
\usepackage{multirow}
\usepackage{mathtools}
\usepackage{amsfonts,amssymb,theorem,mathrsfs}
\usepackage{dsfont}
\usepackage{setspace}
\usepackage{amsmath}
\usepackage{graphicx}
\usepackage[colorlinks,linkcolor=blue,anchorcolor=blue,citecolor=blue]{hyperref}
\usepackage{ulem}
\usepackage{cancel}
\usepackage[justification = raggedright, singlelinecheck = true]{caption}
\usepackage{subcaption}

\begin{document}
\title{Pseudospectrum and (in)stability of black hole total transmission modes}
\preprint{\hfill {\small {ICTS-USTC/PCFT-25-63}}}
\date{\today}

\author{Yu-Sen Zhou}
\email{zhou\_ys@mail.ustc.edu.cn}
\affiliation{Interdisciplinary Center for Theoretical Study and Department of Modern Physics,\\
University of Science and Technology of China, Hefei, Anhui 230026, China}

\author{Ming-Fei Ji}
\email{jimingfei@mail.ustc.edu.cn}

\affiliation{Interdisciplinary Center for Theoretical Study and Department of Modern Physics,\\
University of Science and Technology of China, Hefei, Anhui 230026, China}

\author{Liang-Bi Wu}
\email{liangbi@mail.ustc.edu.cn}
\affiliation{School of Fundamental Physics and Mathematical Sciences, Hangzhou Institute for Advanced Study, UCAS, Hangzhou 310024, China}

\author{Li-Ming Cao}
\email{caolm@ustc.edu.cn}
\affiliation{Interdisciplinary Center for Theoretical Study and Department of Modern Physics,\\
University of Science and Technology of China, Hefei, Anhui 230026, China}
\affiliation{Peng Huanwu Center for Fundamental Theory, Hefei, Anhui 230026, China}

\begin{abstract}
  Total transmission modes (TTMs) are modes with complex frequencies that propagate across a black hole spacetime without reflection. Recently, it is found that suitably tailored time-dependent scattering can excite these complex modes and suppress the reflected signal for the entire duration of the process, a phenomenon referred to as virtual absorption. Motivated by this, we present a study of the spectrum stability of TTMs using pseudospectrum and condition numbers. We focus on perturbations of $d$-dimensional Tangherlini black holes and recast the TTM problem as a generalized eigenvalue problem by utilizing the Eddington-Finkelstein coordinates. The results show that TTMs are generically spectrally unstable, with sensitivity increasing for higher overtones, in close analogy with quasinormal modes. A notable exception is the purely imaginary TTM in the positive imaginary axis in higher dimensions. Its pseudospectrum contours are nearly concentric, and its condition number is orders of magnitude smaller than those of the overtones, indicating enhanced spectral stability. As the spacetime dimension decreases, the condition number grows and becomes much larger in four dimensions in both the energy norm and the $L^2$ norm, suggesting possible spectral instability, although a definitive cross-dimensional conclusion is limited by the lack of a uniform physically preferred norm. Additionally, we confirm that purely imaginary TTMs occur for gravitational vector perturbations, whereas genuinely complex TTM families appear only in sufficiently high dimensions, $d \geqslant 8$, extending earlier claims that placed the onset at $d \geqslant 10$.
\end{abstract}

\maketitle

\section{Introduction}
Linear perturbations of black holes provide a clean arena to study dissipative dynamics in general relativity.  At late times, generic perturbations are dominated by the quasinormal modes (QNMs), which are defined as solutions of the homogeneous perturbation equations that satisfy purely ingoing boundary conditions at the event horizon and purely outgoing conditions at infinity. Their complex frequencies encode the characteristic ringdown signal and depend only on the parameters of the background spacetime; therefore, they play a central role in gravitational wave observations and their precise measurement enables black hole spectroscopy, tests of the Kerr hypothesis, and constraints on possible deviations from general relativity~\cite{Nollert:1999ji, Kokkotas:1999bd, Dreyer:2003bv, Berti:2005ys, Berti:2009kk, Konoplya:2011qq, LIGOScientific:2016aoc, Isi:2019aib, Giesler:2019uxc, Berti:2025hly}.

From the point of view of scattering theory, however, the QNMs are only one prominent class of modes. In addition to the QNMs, which are poles of the reflection and transmission coefficients, there is an additional mode family, known as the total transmission modes (TTMs). At a TTM frequency the reflection coefficient vanishes and an incident wave is transmitted without being reflected through the effective potential barrier~\cite{Chandrasekhar:1985kt, Futterman:1988ni, Andersson:1994tt}. They appear as complex-frequency solutions of the perturbation equation that behave as purely outgoing or purely ingoing plane waves at both asymptotic boundaries. Research has identified TTMs in four dimensions both analytically and numerically~\cite{Berti:2004md, Chen:2025sbz}. For the static Schwarzschild spacetime, Andersson~\cite{Andersson:1994tt} established that the earlier derived algebraically special modes~\cite{Couch:1973zc, Wald:1973wwa, Chandrasekhar:1984mgh} are effectively TTMs. This connection was further elaborated by Maassen van den Brink~\cite{MaassenvandenBrink:2000iwh}, who provided an analytic treatment of gravitational waves at these algebraically special frequencies. For the rotating case, the situation becomes more complicated. Keshet and Neitzke~\cite{Keshet:2007be} performed an asymptotic analysis of QNMs, TTMs, and total-reflection modes. A sequence of studies by Cook \textit{et al.} examined purely imaginary modes and their bifurcation characteristics~\cite{Cook:2014cta, Cook:2016fge, Cook:2016ngj}, and reported a novel class of TTMs whose Schwarzschild limit counterparts appear at complex infinity~\cite{Cook:2018ses, Cook:2022kbb}, which do not satisfy the standard algebraic conditions on the Starobinsky constant~\cite{Starobinskil:1974nkd, Chandrasekhar:1990}. There are also interests in the (quasi)reflectionless modes where the frequency is restricted on the real axis both in gravity theory and in other communities~\cite{Chong:2010, Sweeney2020theory, Rosato:2025byu, Qian:2025occ}.

Similar to QNMs, TTMs are also the characteristic modes of black holes, since they depend only on the parameters of the background black hole. Therefore, TTMs can also serve as a probe of the spacetime and possibly of quantum properties of the black hole~\cite{Keshet:2007be, Kwon:2010mt, Kwon:2011zza}. More recently, it has been shown that suitably tailored initial data can selectively excite a specific TTM in such a way that the whole black hole spacetime effectively acts as a perfect absorber during the scattering~\cite{Tuncer:2025dnp}, a phenomenon known as virtual absorption~\cite{Chong:2010, Baranov:2017}. In this picture, TTMs appear as the characteristic virtual absorption resonances of the scattering problem and thus play a central role in finely controlled black hole scattering experiments. For example, consider a gravitational wave originating from a binary black hole merger or an artificial source scattered by a black hole. If the frequency of this wave matches one of the TTMs of the black hole, the wave will be totally absorbed, i.e., no reflected wave appears. Note that TTMs with a positive imaginary part should also be considered. This differs from the QNM case, where a positive imaginary part indicates dynamical instability. As in the QNM case, the sign of the imaginary part determines both the temporal behavior and the boundary asymptotics of the mode. For QNMs, a positive (negative) imaginary part corresponds to exponential growth (decay) in time and implies that the eigenfunction decays (diverges) exponentially at both boundaries. For TTMs, the temporal role of the imaginary part is the same, but the boundary behavior differs: For the positive imaginary part, the eigenfunction of $\text{TTM}_\text{L}$ ($\text{TTM}_\text{R}$) diverges exponentially at the left (right) boundary, while for the negative imaginary part, the divergence shifts to the opposite boundary. A TTM with a positive imaginary part can be excited by a suitably designed waveform that grows exponentially in time, with the growth rate determined by the imaginary part of the TTM. The reflected wave is then postponed until this exponential growth is eventually terminated by numerical or experimental limitations; see~\cite{Tuncer:2025dnp} for a concrete example.

Given the importance of TTMs in scattering experiments, it is natural to inquire about the robustness of these modes under perturbations induced by the black hole's surrounding environment. It is well known that the QNM problem is an open dissipative problem; thus, its evolution operator is inherently non-Hermitian~\cite{Jaramillo:2020tuu}. Therefore, the spectral stability of QNMs is well assessed by the pseudospectrum. As for the TTM problem, the same mechanism applies: The boundaries admit net energy fluxes and therefore the evolution operator is again non-Hermitian. For these non-Hermitian systems, standard eigenvalue analysis may fail to capture the system's dynamics. It is therefore essential to analyze the pseudospectrum, which captures both eigenvalue sensitivity and potential transient growth, rather than relying solely on the location of isolated eigenvalues~\cite{Trefethen:1999, Trefethen:2005, Jaramillo:2020tuu, Jaramillo:2022kuv, Boyanov:2022ark, Chen:2024mon}. Such pseudospectrum approach has been applied to QNMs of various spacetimes~\cite{Destounis:2021lum, Sarkar:2023rhp, Arean:2023ejh, Courty:2023rxk, Destounis:2023ruj, Cownden:2023dam, Destounis:2023nmb, Boyanov:2023qqf, Cao:2024oud, Carballo:2024kbk, Chen:2024mon, Garcia-Farina:2024pdd, Arean:2024afl, Cao:2024sot, Cai:2025irl, dePaula:2025fqt, Cao:2025qws, Cao:2025afs}, and here we present the first work to extend the pseudospectrum analysis to TTMs. It is known that there are only two pure imaginary TTMs (one for $\text{TTM}_\text{L}$, one for $\text{TTM}_\text{R}$) in four-dimensional Schwarzschild black hole for a fixed $\ell$~\cite{Berti:2004md}. This makes the four-dimensional case less suitable for spectrum (in)stability analyses that are expected to be generic for non-Hermitian open systems. The Tangherlini black hole, which is a simple higher-dimensional generalization of the Schwarzschild black hole, possesses genuinely complex TTMs. We therefore consider perturbations of Tangherlini black holes, recast the TTM problem as a generalized eigenvalue problem on a compact domain, and define pseudospectra with respect to a physically motivated energy norm. We then compute the pseudospectra and condition numbers associated with TTMs. Our main result is that TTMs exhibit spectral instability similar to QNMs, but with a notable exception: There is a spectral stable mode for gravitational perturbation on the imaginary axis.

This paper is organized as follows: In Sec. \ref{sec:setup}, we review the perturbation equations for Tangherlini black holes, define TTMs in terms of their asymptotic behavior, introduce the Eddington-Finkelstein coordinates and compactified radial coordinates, and use the Chebyshev-Lobatto discretization to recast the problem into a generalized eigenvalue problem. Section \ref{sec:pseudo} focuses on the spectral (in)stability of the TTMs using the pseudospectrum and the condition numbers and emphasizes the distinct behaviors of purely imaginary and genuinely complex families. We conclude in Sec. \ref{sec:conclusions} with a summary of our main findings and a discussion of open questions. Technical details concerning the energy norm and alternative definitions of the pseudospectrum are collected in Appendixes \ref{sec:energy_norm} and \ref{sec:pse}, respectively.

\section{Setup}\label{sec:setup}
We consider the Tangherlini black hole, the higher-dimensional generalization of the Schwarzschild black hole in $d$ dimensions~\cite{Tangherlini:1963bw}. Its metric can be written as
\begin{eqnarray}\label{metric}
  \mathrm{d}s^{2}=-f(r)\mathrm{d}t^{2}+f(r)^{-1}\mathrm{d}r^{2} + r^{2}\mathrm{d}\Omega_{d-2}^{2}\, ,
\end{eqnarray}
where $\mathrm{d}\Omega_{d-2}^{2}$ is the line element on the $(d-2)$-dimensional unit sphere,
\begin{eqnarray}\label{metric_function}
  f(r)=1-\bigg(\frac{r_{\text{h}}}{r}\bigg)^{d-3}\, ,
\end{eqnarray}
and $r = r_{\text{h}}$ is the location of the event horizon. The perturbation of this black hole is described by the following master equation~\cite{Kodama:2003jz, Kodama:2003kk}:
\begin{eqnarray}\label{masterEqTime}
  \bigg(\frac{\partial^2}{\partial t^2}-\frac{\partial^2}{\partial x^2}+V\bigg)\Psi=0\, ,
\end{eqnarray}
where the tortoise coordinate $x$ is defined as $\mathrm{d}x=f(r)^{-1}\mathrm{d}r$ and the effective potential,
\begin{eqnarray}
  V=\frac{f}{4r^2}\Bigg\{ 4\ell(\ell+d-3)+(d-2)(d-4)+(1-s^2)(d-2)^2\Big(\frac{r_\text{h}}{r}\Big)^{d-3}\Bigg\} \, ,
\end{eqnarray}
depends on the angular multipole number $\ell$, with $s=0$ describing either massless scalar or gravitational tensor perturbations, $s=2$ corresponds to gravitational vector perturbations, and $s=2/(d-2)$ and $s=2-2/(d-2)$ to the electromagnetic vector and scalar perturbations, respectively~\cite{Kodama:2003jz, Kodama:2003kk, Ishibashi:2003ap, Cardoso:2003vt, Matyjasek:2021xfg, Matyjasek:2024uwo}. Here we focus on the case $s=0$ and $s=2$. While we restrict attention here to the static case, perturbations of higher-dimensional rotating black holes have also been investigated in the literature~\cite{Emparan:2003sy, Kunduri:2006qa, Murata:2008yx, Kodama:2009bf, Dias:2009iu, Dias:2010eu, Dias:2010maa, Lunin:2025yth}. In the frequency domain, we introduce the Fourier ansatz
\begin{eqnarray}\label{fouri}
  \Psi(t,r)=\mathrm{e}^{-\mathrm{i}\omega t}\tilde{\psi}(r)\, ,
\end{eqnarray}
under which the master equation [Eq. \eqref{masterEqTime}] reduces to
\begin{eqnarray}\label{masterEqFreq}
  \bigg(\frac{\mathrm{d}^2}{\mathrm{d}x^2}+\omega^2-V\bigg)\tilde{\psi}=0\, .
\end{eqnarray}
TTMs are defined by their asymptotic behaviors at the boundaries. The right TTM ($\text{TTM}_\text{R}$) and left TTM ($\text{TTM}_\text{L}$) are defined as
\begin{align}
  \text{TTM}_\text{R}\text{:}\quad & \tilde{\psi} \sim \mathrm{e}^{-\mathrm{i}\omega x}\, , \quad \text{as } x \to \pm\infty\, ; \\
  \text{TTM}_\text{L}\text{:}\quad & \tilde{\psi} \sim \mathrm{e}^{+\mathrm{i}\omega x}\, , \quad \text{as } x \to \pm\infty\, .
\end{align}

Following~\cite{Tuncer:2025dnp}, we introduce two new sets of coordinates for the first two dimensions, which are tailored for the TTM problem:
\begin{eqnarray}
  t_\pm=\frac{1}{r_\text{h}}\Big[t\pm x(r)\Big]\, ,\quad\sigma=\frac{r_\text{h}}{r}\, ,
\end{eqnarray}
where the $\pm$ corresponds to the $\text{TTM}_\text{R}$ and the $\text{TTM}_\text{L}$ case, respectively. These sets of coordinates are reminiscent of hyperboloidal coordinates~\cite{Zenginoglu:2011jz, Panossomacedo:2023qzp}, which are, in fact, inspired by these sets of coordinates. One readily finds that $t_+=v/r_\text{h}$ and $t_-=u/r_\text{h}$ are the dimensionless ingoing and outgoing Eddington-Finkelstein coordinates, respectively. Therefore, the boundaries of the compact spatial coordinate, $\sigma=0$ and $\sigma=1$, correspond to different asymptotic regions: For the $\text{TTM}_\text{R}$ case, they correspond to the past null infinity $\mathscr{I}^{-}$ and the future event horizon $\mathscr{H}^{+}$, respectively, whereas for the $\text{TTM}_\text{L}$ case, they correspond to the future null infinity $\mathscr{I}^{+}$ and the past event horizon $\mathscr{H}^{-}$, respectively.

After rescaling the field,
\begin{eqnarray}
  \psi=\mathrm{e}^{\pm\mathrm{i}\omega_\pm x}\tilde{\psi}\, ,
\end{eqnarray}
Eq. (\ref{fouri}) then becomes
\begin{eqnarray}
  \Psi(t_\pm,\sigma)=\mathrm{e}^{\mathrm{i}r_\text{h}\omega_\pm t_\pm}\psi(\sigma)\, ,
\end{eqnarray}
and Eq. (\ref{masterEqFreq}) can be recast as a generalized eigenvalue problem:
\begin{eqnarray}\label{genEigen}
  L_1\psi=\mp \mathrm{i}\omega_\pm r_\text{h}L_2\psi\, ,
\end{eqnarray}
with
\begin{eqnarray}
  L_1=\frac{\mathrm{d}}{\mathrm{d}\sigma}\Big(p\frac{\mathrm{d}}{\mathrm{d}\sigma}\Big)-\frac{r_\text{h}^2V}{p}\, ,\quad L_2=2\frac{\mathrm{d}}{\mathrm{d}\sigma}\, ,
\end{eqnarray}
and
\begin{eqnarray}
  p(\sigma)=-r_\text{h}\frac{\mathrm{d}\sigma}{\mathrm{d}x}=\sigma^2f(r(\sigma))\, .
\end{eqnarray}
The $\mp$ in Eq. (\ref{genEigen}) represents the $\text{TTM}_\text{L}$ and $\text{TTM}_\text{R}$ case, respectively. Because the function $p$ vanishes at the boundaries, TTMs are formulated in terms of the regular solutions $\psi$ of Eq. (\ref{genEigen}). Equation (\ref{genEigen}) is solved numerically by discretizing the differential operators $L_{1}$ and $L_{2}$ into matrices using a Chebyshev-Lobatto grid associated with resolution $N$~\cite{Trefethen:2000, Boyd2013chebyshev, Jaramillo:2020tuu}.

As in the QNM case, spherical symmetry implies that the TTM spectrum is symmetric about the imaginary axis, whereas a reflection about the real axis exchanges $\text{TTM}_\text{R}$ and $\text{TTM}_\text{L}$. Their corresponding pseudospectra share these symmetries. Accordingly, we present only $\text{TTM}_\text{L}$ [i.e., taking the plus sign in Eq. \eqref{genEigen}] and omit its subscript. The TTMs are labeled following the QNM convention, $\omega_0$ is the mode with the largest imaginary part, $\omega_1$ the second largest, and so on. Only modes with non-negative real parts are considered since the modes are symmetric about the imaginary axis. We set $r_\text{h}=1$ in the following presentation.

As reported in~\cite{Tuncer:2025dnp}, a scan over the parameter space $s=0,2$ and $d\geqslant 4$ reveals that purely imaginary TTMs exist if and only if $s=2$, while genuinely complex TTM families arise if and only if $d\geqslant 10$. Our results show some additional details: Genuinely complex TTM families first appear for $d\geqslant 8$. We find genuinely complex TTMs for $d=9,\ell=2$:
\begin{eqnarray}\label{d=9}
  \omega_{s=0}&=&0.99483388-3.45101047\mathrm{i}\, ,\nonumber\\
  \omega_{s=2}&=&0.72889161-2.91253497\mathrm{i}\, ,
\end{eqnarray}
at resolution $N=300$. The absolute values of their differences with their corresponding counterparts under $N=295$ are less than $10^{-18}$ and $10^{-15}$, respectively. There are also genuinely complex TTMs for $d=8,\ell=2$:
\begin{eqnarray}\label{d=8}
  \omega_{s=0}&=&0.20571537-3.08414912\mathrm{i}\, ,\nonumber\\
  \omega_{s=2}&=&0.05-2.56\mathrm{i}\, ,
\end{eqnarray}
at resolution $N=1000$. The absolute values of their differences with their corresponding counterparts under $N=995$ are less than $10^{-9}$ and $10^{-3}$, respectively. The $s=2$ mode appears very close to the imaginary axis; therefore, it is hard to obtain its exact value. Since the results of two resolutions agree only on the first three digits, only these digits are shown. This difficulty is not specific to TTMs. It has long been recognized in the QNM literature that modes near the imaginary axis are numerically delicate to determine, and the same issue arises in a variety of approaches, including Leaver's continued-fraction method, direct integration, and spectral discretizations. The underlying reason is the presence of branch-cut structures near the imaginary axis: For QNMs, the Green function has a branch cut along the negative imaginary axis; for $\text{TTM}_\text{R}$, the reflection amplitude has a branch cut along the positive imaginary axis in four dimensions~\cite{Su:2026fvj}; while for $\text{TTM}_\text{L}$, the corresponding branch cut lies along the negative imaginary axis. In higher dimensions, numerical results also support the persistence of this branch-cut structures, as suggested by the appearance of spurious numerical modes in these regions, in close analogy with the QNM case.

\section{The spectral (in)stability of the total transmission modes}\label{sec:pseudo}
In this section, we analyze the stability of TTMs through the pseudospectrum and the condition number. The TTM problem has been reformulated as a generalized eigenvalue problem [Eq. (\ref{genEigen})]; therefore, a careful treatment of the pseudospectrum is needed.

To begin with, we briefly recall the case of a standard eigenvalue problem, $Ax=\lambda x$. Let $\epsilon>0$. The $\epsilon$ pseudospectrum, $\sigma_{\epsilon}(A)$, of the operator $A$ admits two equivalent definitions. The first, which is more intuitive, is
\begin{eqnarray}\label{def of pseudo 1}
  \sigma_{\epsilon}(A)\coloneq\Big\{\lambda\in\mathbb{C}: \exists\delta A \;\text{with}\; \lVert\delta A\rVert<\epsilon \;\text{such that}\; \lambda\in\sigma(A+\delta A)\Big\}\, ,
\end{eqnarray}
where $\sigma(A)$ denotes the spectrum of $A$.
In computation, the second definition is preferred:
\begin{eqnarray}\label{def of pseudo 2}
  \sigma_\epsilon(A)\coloneq\Big\{\lambda\in\mathbb{C}:\lVert(A-\lambda I)^{-1}\rVert>\epsilon^{-1}\Big\}\, .
\end{eqnarray}
where $I$ represents the identity operator. It is well understood that the choice of norm is crucial~\cite{Gasperin:2021kfv}. We choose the norm in the definitions [(\ref{def of pseudo 1}) and (\ref{def of pseudo 2})] to be the energy norm:
\begin{eqnarray}\label{energy_norm}
  \lVert\psi\rVert_\text{E}\coloneq\sqrt{\langle\psi,\psi\rangle_\text{E}}=\Bigg[\int_{0}^{1}\frac{1}{2}\Big(p\partial_\sigma\bar{\psi}\partial_\sigma\psi+\frac{r_\text{h}^2V}{p}\bar{\psi}\psi\Big)\mathrm{d}\sigma\Bigg]^{\frac{1}{2}}\, ,
\end{eqnarray}
where $\langle\cdot,\cdot\rangle_\text{E}$ is the energy inner product, the bar denotes taking the complex conjugate, and the physical consideration behind this choice can be found in Appendix \ref{sec:energy_norm}. After discretizing the operators and functions into matrices and vectors, the energy inner product and the induced norm are computed using the corresponding Gram matrix $\mathbf{E}$ and its Cholesky factorization $\mathbf{E}=\mathbf{W}^{\star} \cdot\mathbf{W}$~\cite{Jaramillo:2020tuu}. To distinguish continuous operators and functions from their discrete representations, we denote the associated matrices and vectors using boldface:
\begin{eqnarray}\label{E product}
  \langle \mathbf{y},\mathbf{x}\rangle_\text{E}&=&\mathbf{y}^{\star}\mathbf{E}\mathbf{x}\, ,\nonumber\\
  \lVert \mathbf{A}\rVert_\text{E}&=&\lVert \mathbf{W}\cdot\mathbf{A}\cdot\mathbf{W}^{-1}\rVert_\text{2}\, ,\nonumber\\
  \lVert \mathbf{x}\rVert_\text{E}&=&\lVert \mathbf{W}\mathbf{x}\rVert_\text{2}\, ,
\end{eqnarray}
where the $\star$ represents the conjugate transpose and $\lVert\cdot\rVert_2$ represents the $2$-norm. As for the generalized eigenvalue problem, $Ax=\lambda Bx$, however, there are several inequivalent definitions of the $\epsilon$ pseudospectrum, $\sigma_{\epsilon}(A,B)$, and we choose the second definition listed in Appendix~\ref{sec:pse}:
\begin{eqnarray}\label{def of gen pseudo}
  \sigma_{\epsilon}(A,B)&\coloneq&\left\{\lambda\in\mathbb{C}: \exists\delta A \;\text{with}\; \lVert\delta A\rVert<\epsilon \;\text{such that}\; \lambda\in\sigma(A+\delta A,B)\right\}\nonumber\\
  &=&\Big\{\lambda\in\mathbb{C}:\lVert(A-\lambda B)^{-1}\rVert>\epsilon^{-1}\Big\}\, ,
\end{eqnarray}
where $\sigma(A,B)$ is the set of generalized eigenvalues of $(A,B)$. The reason for this choice and a comparison of pseudospectrum with that of the alternative definitions are provided in Appendix~\ref{sec:pse}.

Figure \ref{fig:pse} presents the pseudospectrum of $s=2$ TTMs, with magnified views of $n=0, 1,$ and $2$ modes, and compares them against that of $s=0$ TTMs and $s=2$ QNMs. The contours represent $-\ln\epsilon$ so that higher values correspond to smaller $\epsilon$, with $\epsilon=0$ corresponding to the computed modes. The pseudospectrum of QNMs is computed using definition (\ref{def of pseudo 2}), in which the norm is chosen as the same energy norm in \cite{Jaramillo:2020tuu}. In the lower half-plane, the pseudospectrum of TTMs exhibits behavior similar to that of QNMs. Here $\epsilon$ generally decreases as the imaginary part decreases and tends to zero as it approaches the TTMs. As in the QNM case, the branch cut inherent to asymptotically flat spacetimes introduces spurious numerical modes along the negative imaginary axis. These numerical artifacts can be identified and excluded in practice by comparing results obtained at different grid resolutions. However, since the pseudospectrum is evaluated at a fixed grid resolution, it cannot distinguish these spurious eigenvalues from genuine physical ones. Thus, concentric-circle structures also emerge near the negative imaginary axis, with the spurious modes serving as their centers. In contrast, the pseudospectrum behaves dramatically different in the upper half-plane. In the $s=2$ TTM case, a genuine mode exists on the imaginary axis. Within a considerable region near this mode, the $\epsilon$ contours exhibit an almost concentric circular structure, indicating spectral stability. $\epsilon$ then attains a local maximum near this mode before decreasing with distance from the origin. Conversely, the $s=0$ TTM case lacks this mode and instead displays a local maximum of $\epsilon$ on the imaginary axis that decreases monotonically away from the origin. For the QNM case, contrary to the TTM case, $\epsilon$ increases with the imaginary part, indicating that it is hard to perturb the operator to migrate QNMs into the upper half-plane.
\begin{figure}[htbp]
  \centering
  \begin{subfigure}[t]{0.32\textwidth}
    \includegraphics[width=\linewidth]{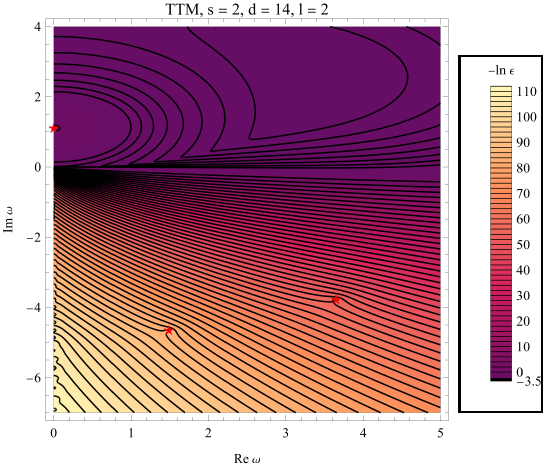}
    \caption{}\label{fig:pse s=2}
  \end{subfigure}\hfill
  \begin{subfigure}[t]{0.32\textwidth}
    \includegraphics[width=\linewidth]{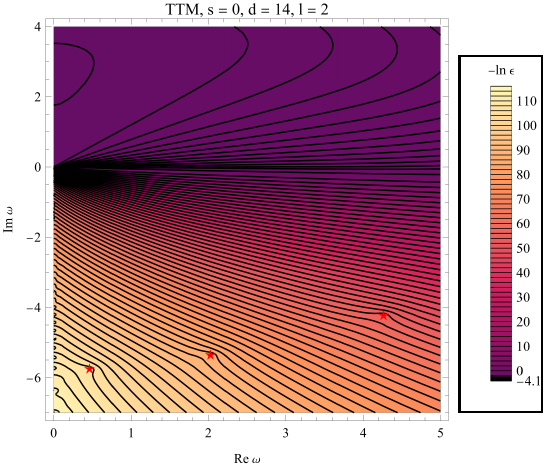}
    \caption{}\label{fig:pse s=0}
  \end{subfigure}\hfill
  \begin{subfigure}[t]{0.32\textwidth}
    \includegraphics[width=\linewidth]{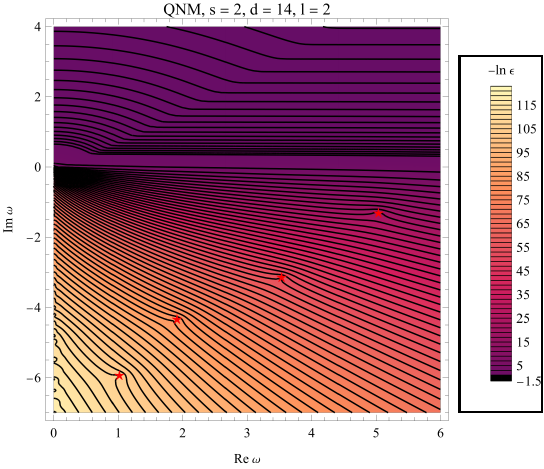}
    \caption{}\label{fig:pse qnm}
  \end{subfigure}\\
  \begin{subfigure}[t]{0.31\textwidth}
    \includegraphics[width=\linewidth]{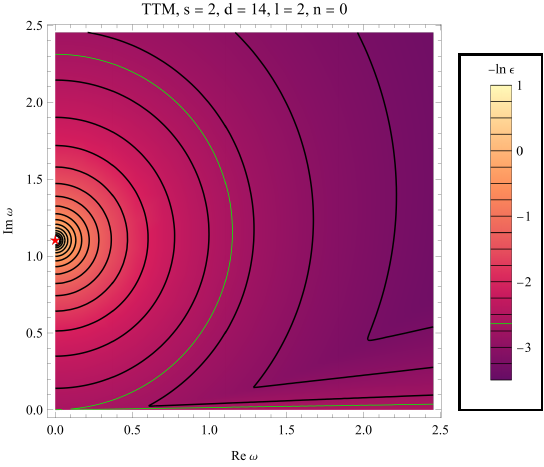}
    \caption{}\label{fig:pse n=0}
  \end{subfigure}\hfill
  \begin{subfigure}[t]{0.32\textwidth}
    \includegraphics[width=\linewidth]{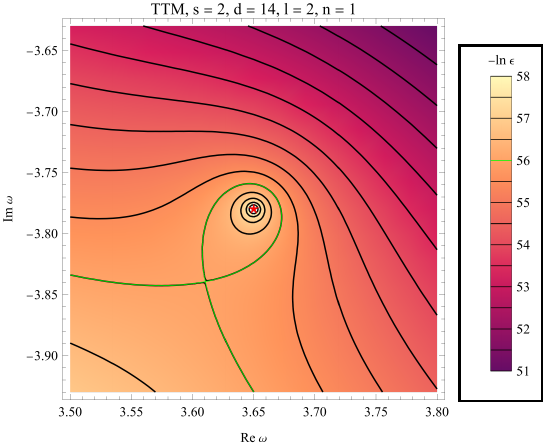}
    \caption{}\label{fig:pse n=1}
  \end{subfigure}\hfill
  \begin{subfigure}[t]{0.32\textwidth}
    \includegraphics[width=\linewidth]{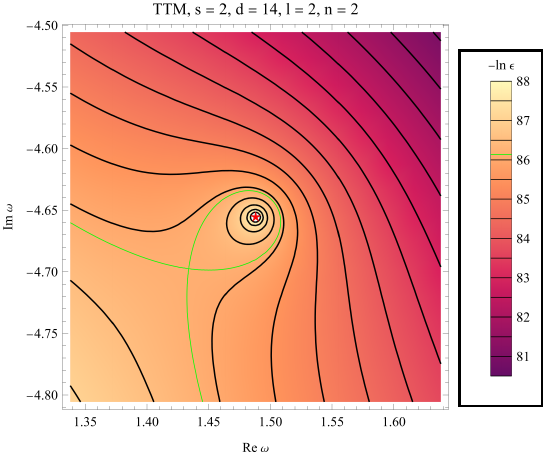}
    \caption{}\label{fig:pse n=2}
  \end{subfigure}
  \caption{Pseudospectra of TTMs with $s=2$ (a) compared with those of TTMs with $s=0$ (b) and QNMs (c). All results are computed with grid resolution $N=200$. The top row displays the overall landscape, while the bottom row enlarges the $n=0, 1,$ and $2$ modes for the $s=2$ case. Red $\star$ symbols indicate the exact TTMs and QNMs, and the green contours mark the transition to an open structure. Due to large variation in the gradient of $-\ln\epsilon$, a specific contour spacing of $1/8$ is used within the following $-\ln\epsilon$ ranges: $[-3.5, -2.5]$ (a), $[-4.065, -2.44]$ (b), and $[-1.5, 1]$ (c).}\label{fig:pse}
\end{figure}

\begin{figure}[htbp]
  \centering
  \includegraphics[width=0.23\linewidth]{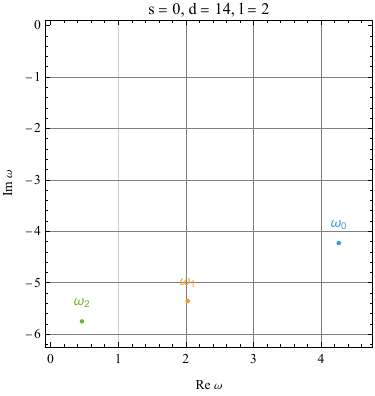}\hfill
  \includegraphics[width=0.23\linewidth]{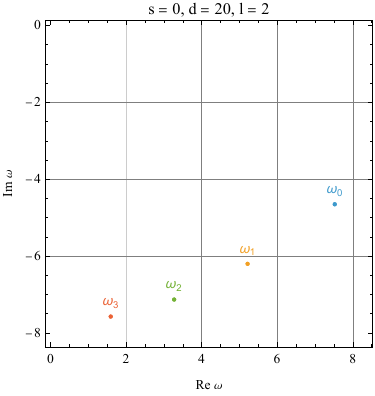}\hfill
  \includegraphics[width=0.23\linewidth]{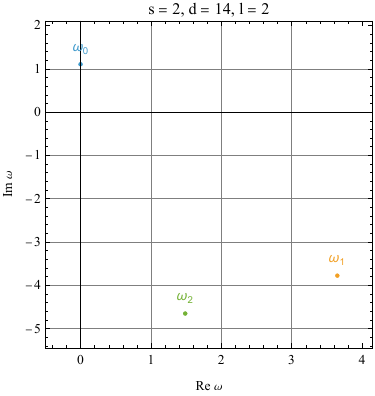}\hfill
  \includegraphics[width=0.23\linewidth]{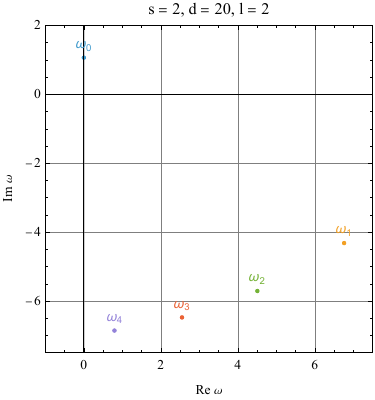}\\
  \includegraphics[width=0.23\linewidth]{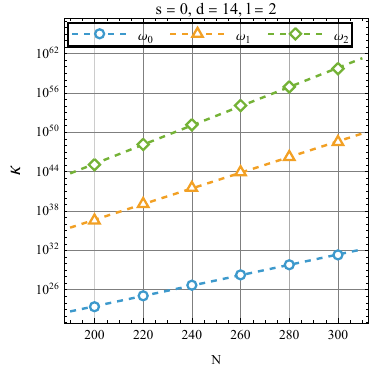}\hfill
  \includegraphics[width=0.23\linewidth]{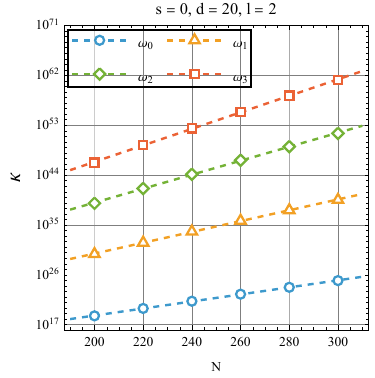}\hfill
  \includegraphics[width=0.23\linewidth]{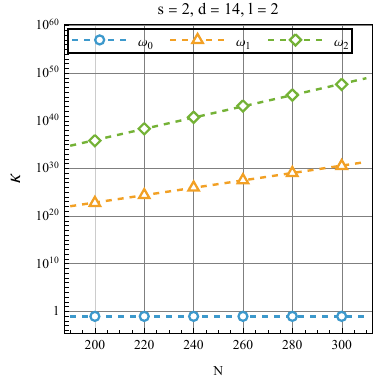}\hfill
  \includegraphics[width=0.23\linewidth]{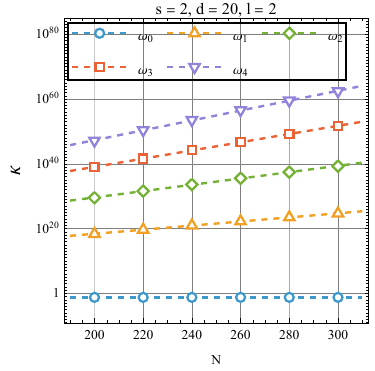}
  \caption{The TTMs $\omega_n$ (top panels) and their condition numbers $\kappa$ (bottom panels) as functions of $N$ for $s=0,2$ and $d=14,20$ with $\ell=2$ obtained within the resolution $N=300$. Dashed lines show linear fits to the condition numbers which align with the numerical values very accurately.}
  \label{fig:cn}
\end{figure}

We introduce another quantity, namely the condition number, to measure the spectral stability of an eigenvalue. For a generalized eigenvalue problem $\mathbf{A}\mathbf{x}=\lambda\mathbf{B}\mathbf{x}$ and a perturbation $(\Delta \mathbf{A},\Delta \mathbf{B})$ on $(\mathbf{A}, \mathbf{B})$, the eigenvalue changes as
\begin{eqnarray}
  \Delta \lambda=\frac{\langle \mathbf{y}_\text{E},(\Delta \mathbf{A}-\lambda\Delta \mathbf{B})\mathbf{x}\rangle_\text{E}}{\langle \mathbf{y}_\text{E},\mathbf{B}\mathbf{x}\rangle_\text{E}}+o(\lVert\Delta \mathbf{A}\rVert_\text{E},\lVert\Delta \mathbf{B}\rVert_\text{E})\, ,
\end{eqnarray}
where $\mathbf{x}$ is the right generalized eigenvector associated with $\lambda$, $\langle\cdot,\cdot\rangle_\text{E}$ represents the energy inner product (see Appendix \ref{sec:energy_norm}), and $\mathbf{y}_\text{E}$ is the left generalized eigenvector with respect to the energy inner product, which satisfies
\begin{eqnarray}
  (\mathbf{A}-\lambda \mathbf{B})^\dagger \mathbf{y}_\text{E}=\mathbf{E}^{-1}(\mathbf{A}-\lambda \mathbf{B})^\star \mathbf{E}\mathbf{y}_\text{E}=0\, ,
\end{eqnarray}
in which a dagger represents the adjoint with respect to the energy inner product, and $\mathbf{E}$ is the corresponding Gram matrix of energy norm.
Consider the case $\Delta \mathbf{B}=0$ and $\lVert\Delta \mathbf{A}\rVert<\epsilon$. The condition number can be defined as
\begin{eqnarray}\label{condition_number}
  \kappa(\lambda)\coloneq\lim_{\epsilon\rightarrow0}\sup_{\lVert\Delta \mathbf{A}\rVert_\text{E}<\epsilon}\frac{\lvert\Delta \lambda\rvert}{\epsilon}=\frac{\lVert \mathbf{y}_\text{E}\rVert_\text{E}\lVert \mathbf{x}\rVert_\text{E}}{\lvert\langle \mathbf{y}_\text{E},\mathbf{B}\mathbf{x}\rangle_\text{E}\rvert}\,,
\end{eqnarray}
in which the energy inner product and norm are calculated using Eqs. (\ref{E product}). Therefore, a mode with a larger condition number is more spectrally unstable. Unlike the standard eigenvalue problem, where the condition number is bounded below by $1$, in our case, it can fall below $1$. For example, if we multiply the generalized eigenvalue problem by a constant $a$ on both sides, the eigenvalues and their corresponding eigenvectors remain unchanged, but the condition numbers are scaled by a factor of $1/a$:
\begin{eqnarray}\label{amb in cn}
  \mathbf{A}\mathbf{x}=\lambda\mathbf{B}\mathbf{x}&\longrightarrow& (a\mathbf{A})\mathbf{x}=\lambda(a\mathbf{B})\mathbf{x}\, ,\nonumber\\
  \kappa&\longrightarrow& \kappa/a\, .
\end{eqnarray}
Therefore, the condition number of an eigenvalue by itself is meaningless; only the relative ratios of condition numbers between different eigenvalues are of importance. Hereafter, we fix $L_2=2\mathrm{d}/\mathrm{d}\sigma$ to mitigate this ambiguity. We show the TTMs for several parameters and compute their condition numbers for a range of grid resolutions, as shown in Fig. \ref{fig:cn}. At fixed grid resolution, the condition number increases with the overtone index, so higher overtones are more ill conditioned and therefore more spectrally unstable. For each TTM, the condition number also grows as the grid is refined, and this growth is steeper for higher overtones. Remarkably, for $s=2$ the fundamental modes behave differently. Their condition numbers converge to small values as the grid is refined. Specifically, if one defines
\begin{eqnarray}\label{Delta_kappa}
  \Delta\kappa=\frac{\kappa^{(N_2)}-\kappa^{(N_1)}}{N_2-N_1}\, ,
\end{eqnarray}
then for the fundamental modes in $s=2$, $\Delta\kappa\approx2.48\times10^{-12}$ for $d=14$ and $\Delta\kappa\approx1.67\times10^{-13}$ for $d=20$ with $N_1=200$ and $N_2=300$. This small condition number sharply contrasts with the overtones, whose condition numbers are large and increase rapidly under grid refinement. Such robustness indicates the spectral stability of the fundamental mode for $s=2$.

Condition numbers are closely related to the pseudospectrum, especially since we adopt a definition compatible with that of the pseudospectrum. In particular, the condition number of a given mode characterizes the local behavior of the pseudospectrum near that mode. Upon careful inspection of the definition of the condition number (\ref{condition_number}), it follows that, for fixed $\epsilon$, the supremum of $\lvert\Delta\lambda\rvert$ corresponds to the distance of the farthest point of the $\epsilon$-pseudospectrum contour from the eigenvalue. Hence, as the increase of the grid resolution, the convergence of the condition number indicates the convergence of the pseudospectrum in the vicinity of that mode. The fact that the pseudospectrum converges in the upper half-plane but not in the lower half-plane suggests the presence of an underlying regularity-dependent structure. In asymptotically anti-de Sitter spacetimes, this convergence is organized into horizontal regions of the complex-frequency plane determined by the regularity built into the norm, with access to more strongly damped regions requiring stronger regularity control~\cite{Warnick:2013hba,Boyanov:2023qqf,Arean:2024afl}. An analogous regularity dependence also appears in the de Sitter case, where convergence is likewise restricted to the region compatible with the chosen level of regularity~\cite{Warnick:2024usx}. For asymptotically flat spacetimes, however, and in particular for TTMs, both the appropriate choice of norm and the proper treatment of the branch cut remain less clear: The former may need to incorporate the behavior at null infinity, possibly through a Gevrey-type norm as in~\cite{Gajic:2019qdd,Gajic:2019oem,Gevrey:22021}, while the latter requires a separate careful analysis. We leave these issues for future work.

\begin{figure}[htbp]
  \centering
  \begin{subfigure}[t]{0.25\textwidth}
    \includegraphics[width=\linewidth]{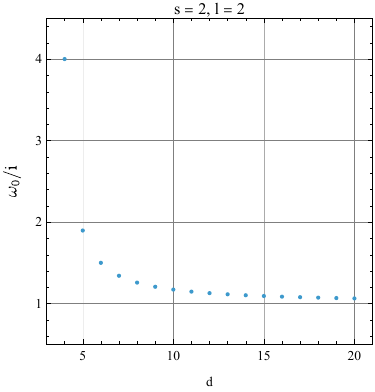}
    \caption{}\label{fig:4da}
  \end{subfigure}
  \hfill
  \begin{subfigure}[t]{.35\textwidth}
    \includegraphics[width=\linewidth]{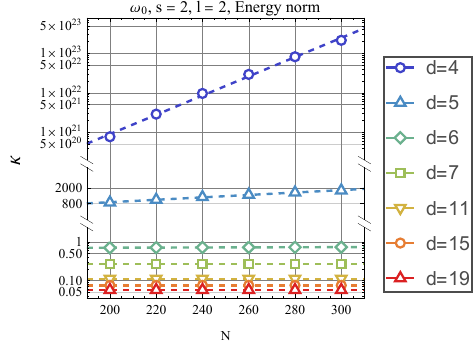}
    \caption{}\label{fig:4db}
  \end{subfigure}
  \begin{subfigure}[t]{.35\textwidth}
    \includegraphics[width=\linewidth]{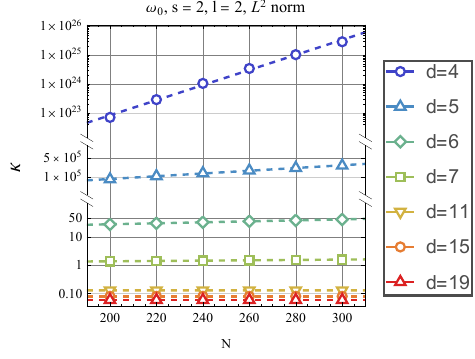}
    \caption{}\label{fig:4dc}
  \end{subfigure}
  \caption{(a) The imaginary part of the modes in positive imaginary axis for several spacetime dimensions. (b) Condition numbers of the modes in positive imaginary axis for several spacetime dimensions and numerical grid resolutions using energy norm. (c) The same as (b), but the condition number now adopts $L^2$ inner product and norm. The vertical axes are folded twice, but the scale remains consistent throughout.}
  \label{fig:4d}
\end{figure}

We now comment on the four-dimensional case, where each sector contains only a single mode, namely one $\text{TTM}_\text{L}$ mode and one $\text{TTM}_\text{R}$ mode. Owing to the ambiguity inherent in the generalized eigenvalue problem, only comparisons of condition numbers and pseudospectra between different modes under the same set of parameters are directly meaningful. At first sight, this appears to prevent even a relative assessment of stability. However, since the operator $L_2=2\mathrm{d}/\mathrm{d}\sigma$ in Eq. (\ref{genEigen}) is independent of the spacetime dimension $d$, the condition number is not affected by the ambiguity discussed in Eq.~(\ref{amb in cn}). Therefore, a qualitative comparison across different dimensions should still be meaningful, although it does not constitute a rigorous proof. Focusing on the modes in the positive imaginary axis, we computed their condition numbers for several spacetime dimensions, as shown in Fig.~\ref{fig:4db}. Note that the energy norm depends on the spacetime dimensions, so we also provide a supplementary $L^2$-norm result in Fig.~\ref{fig:4dc}, which is independent of the spacetime dimensions. The $L^2$-norm condition number is obtained by replacing the energy inner product and energy norm in the definition of the condition number~(\ref{condition_number}) with $L^2$ inner product and norm. The definition of $L^2$ inner product and norm can be found at the end of Appendix \ref{sec:energy_norm}. The results show that the condition number in both norm increases as $d$ decreases and becomes much larger in four dimensions, suggesting possible spectral instability. However, since a uniform physically preferred norm among different dimensions is not available, this cross-dimensional comparison should be interpreted as diagnostic evidence rather than a definitive conclusion. This behavior cannot be understood solely from the location of the mode in the complex-frequency plane. As shown in Fig.~\ref{fig:4da}, the purely imaginary fundamental mode moves upward along the positive imaginary axis as $d$ decreases. If the regularity structure were the same for all dimensions, this upward motion alone would not suggest worse convergence. Indeed, at fixed $d$, comparison among different overtones indicates that the convergence of the pseudospectrum becomes worse for modes with smaller imaginary parts, in analogy with the regularity-dependent structure seen in the anti-de Sitter case~\cite{Warnick:2013hba,Boyanov:2023qqf,Arean:2024afl}. The observed dimensional trend is therefore opposite to what one would expect from the mode location alone. This suggests that varying $d$ changes the regularity structure of the pseudospectrum in a dramatic way.

\section{Conclusions and discussion}\label{sec:conclusions}
In this work, we present the first study of the spectrum (in)stability of TTMs using pseudospectra and eigenvalue condition numbers. Motivated by the recent realization that TTMs can be selectively excited by tailored, initial data to produce total absorption in black-hole scattering~\cite{Tuncer:2025dnp}, we addressed a basic but largely unexplored question: How robust are TTMs as eigenvalues of a non-Hermitian evolution operator?

Our analysis focuses on linear perturbations of $d$-dimensional Tangherlini black holes. By adopting Eddington-Finkelstein coordinates, we recast the TTM problem into the generalized eigenvalue problem Eq. (\ref{genEigen}). This formulation allows a direct pseudospectral analysis after Chebyshev-Lobatto discretization. The energy inner product and associated norm are used to compute the pseudospectrum and the eigenvalue condition number.

Two main conclusions emerge. First, TTMs are generically strongly spectrally unstable, closely paralleling what is by now well established for QNMs~\cite{Jaramillo:2020tuu}. In our computations, the pseudospectra around most TTMs quickly transition to an open structure, and the corresponding $\epsilon_c$ values that mark this transition decrease significantly as the overtone number increases. The corresponding condition numbers are several orders of magnitude larger than that of the fundamental mode in the higher-dimensional $s=2$ case, and they increase rapidly as the overtone number rises. This behavior indicates that small perturbations of the evolution operator (for instance, induced by modifications of the effective potential) can lead to large shifts of the TTMs and that this sensitivity becomes more severe for higher overtones. Second, we identified a notable exception: For gravitational vector perturbations ($s=2$) in higher dimensions, there exists a purely imaginary TTM whose pseudospectral contours are nearly concentric over a sizable neighborhood, and whose condition number is orders of magnitude smaller than those of the overtones. This points to an enhanced spectral stability of this specific mode, in sharp contrast with the behavior of the rest of the TTM spectrum. Inspecting this mode across several spacetime dimensions shows that its condition number increases as $d$ decreases and becomes much larger in four dimensions, both in the energy norm and in the $L^2$ norm. This suggests possible spectral instability in lower dimensions; however, because there is no uniform physically preferred norm for comparing different dimensions, this should be regarded as diagnostic evidence rather than a definitive conclusion. Moreover, as we adopt a definition compatible with that of the pseudospectrum, the convergence of the condition number indicates that the pseudospectrum converges in the vicinity of that mode. Thus, the results suggest that the pseudospectrum converges in the upper half-plane in higher dimensions, and we suspect that a similar regularity-dependent structure may be present, analogous to that discussed in~\cite{Warnick:2013hba,Boyanov:2023qqf,Arean:2024afl,Warnick:2024usx,Gajic:2019qdd,Gajic:2019oem,Gevrey:22021}. In addition, we confirm that this purely imaginary TTM occurs only for $s=2$, while genuinely complex TTM families arise only in sufficiently high dimensions. Our numerical results indicate that such complex families can appear already by $d=8$ [cf. (\ref{d=8}) and (\ref{d=9})], extending earlier claims that they do not appear until $d\geqslant 10$~\cite{Tuncer:2025dnp}.

These findings have several implications for controlled scattering scenarios, since virtual absorption relies on precise time-domain tuning to trigger a specific mode~\cite{Tuncer:2025dnp}. Strong instability suggests that the relevant TTMs may be highly susceptible to environmental perturbations (e.g., external matter distributions or model uncertainties), particularly for higher overtones. Conversely, the existence of an unusually spectrally stable, purely imaginary TTM suggests a potentially more robust target for detection. Clarifying whether this stability is accidental, symmetry protected, or tied to a deeper structure (for example, whether it is ensured by the vanishing real part, or originates from the reversed sign of the imaginary part relative to conventional modes) is an immediate open question.

It is possible to formulate the problem of solving TTMs into a standard eigenvalue problem instead of a generalized eigenvalue problem by adopting other spacetime foliation while still maintaining the current advantage that the only requirement on the solutions is regularity. Specifically, one may introduce a time coordinate $\tau$ such that the constant-$\tau$ slices define the hypersurfaces of the foliation. These hypersurfaces are chosen to be spacelike in the bulk yet approach null slices near the boundaries. Taking $\text{TTM}_\text{L}$ as an example, the foliation should be carefully constructed so that it becomes asymptotically outgoing null both at the past event horizon and at future null infinity. Consequently, it naturally adopts to the corresponding boundary conditions at these boundaries. The eigenvalue problem then reduces to solving regular solutions of the resulting equation. Meanwhile, since the foliation only becomes asymptotically null at the boundaries, the resulting equation contains second-order time derivatives with a coefficient which vanishes only at the boundaries. Thus, by introduction a new variable $\Pi=\partial_\tau\Psi$, the resulting equation can be reduced to first order in $\tau$, yielding a standard eigenvalue problem. In the framework of the standard eigenvalue, both the notion of the stability and the definition of the pseudospectrum become less ambiguous. Nonetheless, we believe our conclusions are qualitatively robust.

It is natural to extend the present work in the following directions: the stability of the TTMs of rotating backgrounds, given that Kerr TTMs have richer structures; how specific perturbations deform the TTM spectrum; and the implications of this spectrally stable TTM for time-domain dynamics. Overall, our results place TTMs on similar conceptual footing to QNMs regarding non-normality and spectral instability, while revealing an intriguing exception that may be particularly relevant for precision scattering and absorption experiments.

\section*{ACKNOWLEDGMENT}
We thank the referee for inspiring comments regarding the convergence of condition numbers and for inquiring other possible coordinates. This work is supported in part by the National Key R\&D Program of China Grant No. 2022YFC2204603 and by the National Natural Science Foundation of China with Grants No. 12475063, No. 12075232 and No. 12247103. This work is also supported by the National Natural Science Foundation of China with Grants No. 12505067.

\appendix

\section{THE ENERGY NORM}\label{sec:energy_norm}
In this appendix, we outline the derivation of the physically relevant energy norm used to compute the pseudospectrum of TTMs~\cite{Gasperin:2021kfv}. For the QNM case, we adopt the hyperboloidal coordinate constructed by out-in strategy~\cite{Panossomacedo:2023qzp} and utilize the energy norm defined in~\cite{Jaramillo:2020tuu}. The master equation~[Eq. (\ref{masterEqTime})] can be written equivalently as
\begin{eqnarray}\label{effMasterEq}
  \eta^{ab}\mathring{\nabla}_a\mathring{\nabla}_b\Psi-V\Psi=0\, ,
\end{eqnarray}
where $\eta_{ab}$ is the metric of a $(1+1)$-dimensional Minkowski spacetime, and $\mathring{\nabla}$ denotes the covariant derivative associated with $\eta_{ab}$. Equation (\ref{effMasterEq}) can be regarded as originating from the effective action
\begin{eqnarray}\label{effAction}
  S=-\int\mathrm{d}^2y\sqrt{-\eta}\frac{1}{2}\Big(\eta^{ab}\mathring{\nabla}_a\bar{\Psi}\mathring{\nabla}_b\Psi+V\bar{\Psi}\Psi \Big)\, ,
\end{eqnarray}
which is associated with a stress-energy tensor,
\begin{eqnarray}
  T_{ab}=\frac{1}{2}\Big(\mathring{\nabla}_a\bar{\Psi}\mathring{\nabla}_b\Psi+\mathring{\nabla}_a\Psi\mathring{\nabla}_b\bar{\Psi}\Big)-\frac{1}{2}\eta_{ab}\Big(\eta^{cd}\mathring{\nabla}_c\bar{\Psi}\mathring{\nabla}_d\Psi+V\bar{\Psi}\Psi\Big)\, .
\end{eqnarray}
Accordingly, one can evaluate the conserved energy associated with time translations generated by the Killing vector $\xi^a$ on the constant-$t_\pm$ hypersurface $\Sigma_{t_\pm}$ as
\begin{eqnarray}
  E&=&\int_{\Sigma_{t_\pm}}T_{ab}\xi^a n^b\mathrm{d}r\nonumber\\
  &=&\int_{r_\text{h}}^{+\infty}\frac{1}{2}r_\text{h}\Big(f\partial_r \bar{\Psi}\partial_r \Psi+\frac{V}{f}\bar{\Psi}\Psi\Big)\mathrm{d}r\nonumber\\
  &=&\int_{0}^{1}\frac{1}{2}\Big(p\partial_\sigma\bar{\Psi}\partial_\sigma\Psi+\frac{r_\text{h}^2V}{p}\bar{\Psi}\Psi\Big)\mathrm{d}\sigma\, ,
\end{eqnarray}
where coordinate basis vector fields are chart dependent. To avoid ambiguity, we denote the $r$-direction basis vector in the $\{t_\pm,r\}$ chart by $\bigg(\dfrac{\partial}{\partial r}\bigg)^a_{t_\pm}$. It is found that
\begin{equation*}
  \left(\frac{\partial}{\partial r}\right)^a_{t_\pm}=\left(\frac{\partial}{\partial r}\right)^a_{t}+\left(\frac{\partial t}{\partial r}\right)_{t_\pm}\left(\frac{\partial}{\partial t}\right)^a_{r}=\left(\frac{\partial}{\partial r}\right)^a_{t}\mp\frac{1}{f}\left(\frac{\partial}{\partial t}\right)^a_{r}\neq\left(\frac{\partial}{\partial r}\right)^a_{t}\, .
\end{equation*}
Since no such ambiguity arises for $\xi^a$, we omit its subscript.
\begin{eqnarray}
  \xi^a=\bigg(\frac{\partial}{\partial t_\pm}\bigg)^a\, ,\quad n^a=-r_\text{h}g^{ab}(\mathrm{d}t_\pm)_b=\mp\bigg(\frac{\partial}{\partial r}\bigg)^a_{t_\pm}\, ,
\end{eqnarray}
and $n^a$ is the vector normal to $\Sigma_{t_\pm}$.
The physically relevant energy inner product is inspired by the energy on the $t_\pm=0$ hypersurface as
\begin{eqnarray}\label{energyInnerProd}
  \langle\psi_1,\psi_2\rangle_\text{E}=\int_{0}^{1}\frac{1}{2}\Big(p\partial_\sigma\bar{\psi}_1\partial_\sigma\psi_2+\frac{r_\text{h}^2V}{p}\bar{\psi}_1\psi_2\Big)\mathrm{d}\sigma\, ,
\end{eqnarray}
which is positive definite; thus, the energy norm can be defined closely as
\begin{eqnarray}
  \lVert\psi\rVert_\text{E}=\sqrt{\langle\psi,\psi\rangle_\text{E}}\, .
\end{eqnarray}
The energy norm has a clear physical interpretation, but it inevitably depends on the spacetime dimension. To provide a uniform baseline across different dimensions, we also introduce the $L^2$ inner product and $L^2$ norm:
\begin{eqnarray}
  \langle\psi_1,\psi_2\rangle_2&=&\frac{1}{2}\int_{0}^{1}\bar{\psi}_1\psi_2\mathrm{d}\sigma\,,\\
  \lVert\psi\rVert_2&=&\sqrt{\langle\psi,\psi\rangle_2}\,.
\end{eqnarray}
Since the $L^2$ norm does not have a direct physical interpretation in the present context, results based on this norm should be regarded as a mathematical reference rather than a definitive physical criterion.

\section{THE DEFINITIONS OF PSEUDOSPECTRUM FOR A GENERALIZED EIGENVALUE PROBLEM}\label{sec:pse}
\begin{figure}[htbp]
  \centering
  \begin{subfigure}[t]{0.48\textwidth}
    \includegraphics[width=\linewidth]{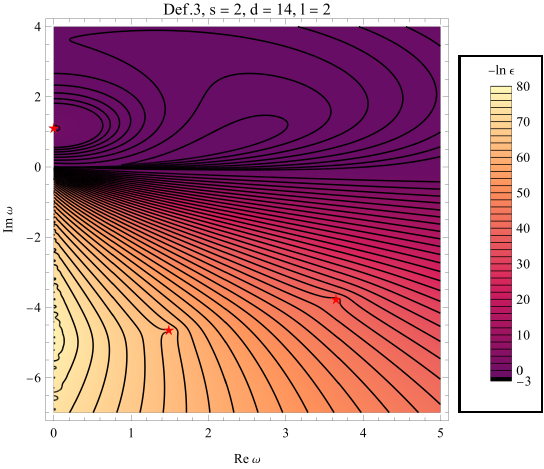}
    \caption{}\label{fig:def3 pse s=2}
  \end{subfigure}\hfill
  \begin{subfigure}[t]{0.48\textwidth}
    \includegraphics[width=\linewidth]{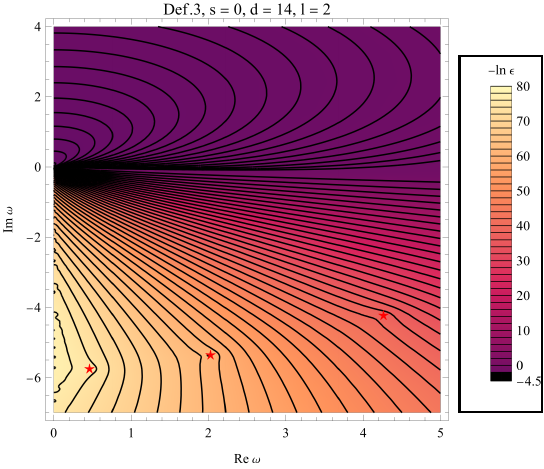}
    \caption{}\label{fig:def3 pse s=0}
  \end{subfigure}\\
  \begin{subfigure}[t]{0.30\textwidth}
    \includegraphics[width=\linewidth]{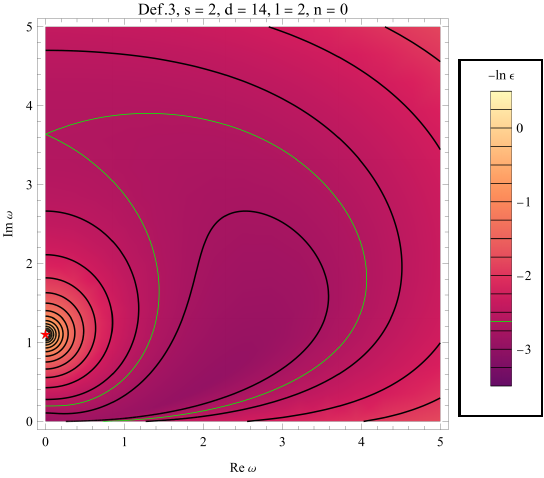}
    \caption{}\label{fig:def3 pse n=0}
  \end{subfigure}\hfill
  \begin{subfigure}[t]{0.32\textwidth}
    \includegraphics[width=\linewidth]{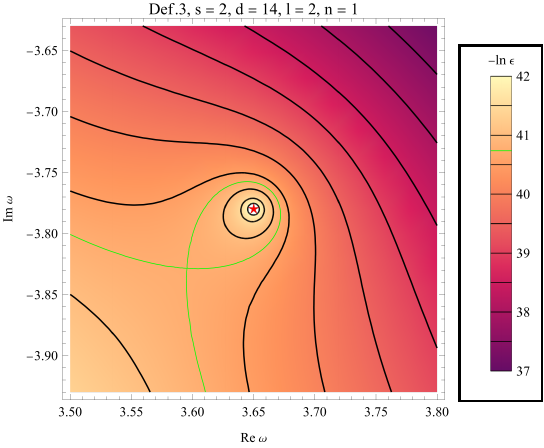}
    \caption{}\label{fig:def3 pse n=1}
  \end{subfigure}\hfill
  \begin{subfigure}[t]{0.32\textwidth}
    \includegraphics[width=\linewidth]{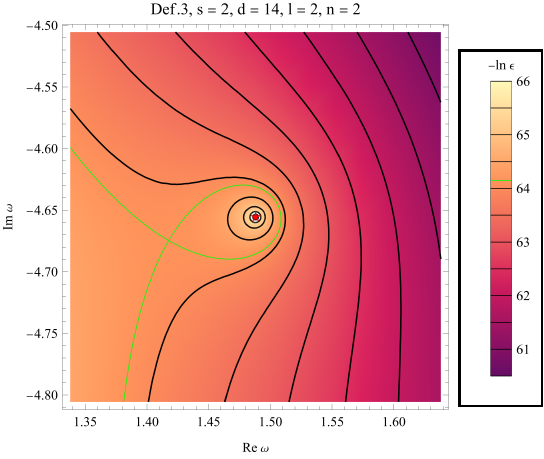}
    \caption{}\label{fig:def3 pse n=2}
  \end{subfigure}
  \caption{Pseudospectra of TTMs adopting the third definition (\ref{def3}) of a generalized eigenvalue problem. All the parameters are the same as the Fig. \ref{fig:pse}. The top row displays the overall landscape, while the bottom row enlarges the $n=0, 1,$ and $2$ modes for the $s=2$ case. Red $\star$ symbols indicate the exact TTMs, and the green contours mark the transition to an open structure. Due to large variation in the gradient of $-\ln\epsilon$, a specific contour spacing of $1/8$ is used within the following $-\ln\epsilon$ ranges: $[-3, -2]$ (a) and $[-4.5, -2]$ (b).}\label{fig:pse def 3}
\end{figure}

In this appendix, we show the consideration behind the choice of the specific definition of the pseudospectrum adopted in the text. Given a generalized eigenvalue problem $Ax=\lambda Bx $, where $A$ and $B$ are two operators and $x$ is the generalized eigenvector associated with the generalized eigenvalue $\lambda$, unlike the standard eigenvalue problem, its pseudospectrum has several inequivalent definitions in the literature, as summarized in~\cite{Trefethen:2005}.
\begin{description}
  \item[\textbf{Definition B1}]
    \begin{eqnarray}\label{def1}
      \sigma^{(1)}_{\epsilon}(A,B)\coloneq\sigma_{\epsilon}(B^{-1}A)\, ,
    \end{eqnarray}
    \textit{which converts the generalized eigenvalue problem to a standard eigenvalue problem. This requires $B$ to be nonsingular.}
  \item[\textbf{Definition B2}]
    \begin{eqnarray}\label{def2}
      \sigma^{(2)}_{\epsilon}(A,B)&\coloneq&\left\{\lambda\in\mathbb{C}: \exists\delta A \;\text{with}\; \lVert\delta A\rVert<\epsilon \;\text{such that}\; \lambda\in\sigma(A+\delta A,B)\right\}\nonumber\\
      &=&\Big\{\lambda\in\mathbb{C}:\lVert(A-\lambda B)^{-1}\rVert>\epsilon^{-1}\Big\}\,,
    \end{eqnarray}
    \textit{which is equivalent to considering the perturbations on $A$ solely.}
  \item[\textbf{Definition B3}]
    \begin{eqnarray}\label{def3}
      \sigma^{(3)}_{\epsilon;\alpha,\beta}(A,B)
      &\coloneq& \left\{\lambda\in\mathbb{C}:\exists\delta A,\delta B\;\text{with}\;\lVert\delta A\rVert\le\alpha\epsilon,\lVert\delta B\rVert\le\beta\epsilon\;\text{such that}\; \lambda\in\sigma(A+\delta A,B+\delta B)\right\}\nonumber\\
      &=&\left\{\lambda\in\mathbb{C}:\ \|(A-\lambda B)^{-1}\|>\frac{1}{\epsilon(\alpha+|\lambda|\beta)}\right\}\, ,
    \end{eqnarray}
    \textit{where $\alpha,\beta>0$ are fixed weights.}
  \item[\textbf{Definition B4}]
    \begin{eqnarray}\label{def4}
      \sigma^{(4)}_{\epsilon}(A,B)&\coloneq&\left\{\lambda\in\mathbb{C}:\exists\delta A\;\text{with}\; \lVert\delta A\rVert_\text{B}<\epsilon \;\text{such that}\;\lambda\in\sigma(A+\Delta A,\ B)\right\}\nonumber\\
      &=&\Big\{\lambda\in\mathbb{C}:\lVert(A-\lambda B)^{-1}\rVert_\text{B}>\epsilon^{-1}\Big\}\, .
    \end{eqnarray}
    \textit{where it is assumed that $B$ is a Hermitian positive-definite operator, and $\lVert\cdot\rVert_\text{B}$ denotes the norm induced by $B$. Specializing the norm in Definition \ref{def2} to $\lVert\cdot\rVert_\text{B}$ yields Definition \ref{def4}.}
\end{description}
For a clearer manifestation of non-normality (e.g., the transient effect), the first definition (\ref{def1}) would be most suitable and is likewise used in~\cite{Chen:2024mon}. In our setting, however, the operator $B$, corresponding to our $L_2$ in Eq. (\ref{genEigen}), is not invertible, rendering Definition \ref{def1} inapplicable. From the standpoint of a pure mathematical problem, the third definition (\ref{def3}) would best illustrates eigenvalue perturbations. Yet in our problem $L_2=2\mathrm{d}/\mathrm{d}\sigma$ is a differential operator that should remain fixed under physical perturbations (e.g., perturbations on the effective potential or the background spacetime), so we do not adopt Definition \ref{def3}. The fourth definition (\ref{def4}) effectively considers perturbations of $A$ alone and measures them in the norm induced by $B$. Because here $L_2$ is neither positive-definite nor Hermitian and thus cannot induce a norm, Definition \ref{def4} is likewise inapplicable. Therefore the second definition (\ref{def2}), which is also used in~\cite{Boyanov:2023qqf, Cownden:2023dam}, is adopted. We specialize the norm as the energy norm and henceforth omit the superscript in $\sigma^{(2)}_{\epsilon}(A,B)$. For completeness, Fig. \ref{fig:pse def 3} presents results of the pseudospectrum corresponding to Fig. \ref{fig:pse} under the same parameters but using the third definition (\ref{def3}). The weights therein are chosen as $\alpha = \lVert L_1\rVert_\text{E}/(\lVert L_1\rVert_\text{E}+\lVert L_2\rVert_\text{E})$ and $\beta = \lVert L_2\rVert_\text{E}/(\lVert L_1\rVert_\text{E}+\lVert L_2\rVert_\text{E})$. Unexpectedly, $\alpha \approx 0.912$ and $\beta \approx 0.088$ remain almost identical for both $s=0$ and $s=2$; deviations only begin at the fourth decimal place.

\bibliography{mainRef}
\bibliographystyle{apsrev4-1}

\end{document}